\documentclass[11pt]{article}
\usepackage{epsfig}
\begin{document}
\def\be{\begin{equation}}
\def\ee{\end{equation}}
                         \def\bearr{\begin{eqnarray}}
                         \def\eearr{\end{eqnarray}}
\def\benum{\begin{enumerate}}
\def\eenum{\end{enumerate}}
\def\bitem{\begin{itemize}}
\def\eitem{\end{itemize}}
                         \def\eg{ {\em e.g.}~}
                         \def\etal{ {\em et al.}~}
                         \def\ie{ {\em i.e.}~}
                         \def\viz{ {\em viz.}~}

\def\go{\rightarrow}
\def\goes{\longrightarrow}
\def\hrar{\hookrightarrow}
\def\bul{\bullet}
\def\mET{E_T \hspace{-1.1em}/\;\:}
\def\mpT{p_T \hspace{-1em}/\;\:}
\def\rpv{$R_p \hspace{-1em}/\;\:$}
\def\cpv{$\; \not \!\!\!\!CP $}
\def\lv{$\; \not \!\! L$}
\def\bv{$\; \not \!\! B$}
\def\rp{$R$-parity}
\def\tb{\tan \beta}
                         \def\N0{\widetilde \chi^0}
                         \def\Cp{\widetilde \chi^+}
                         \def\Cm{\widetilde \chi^-}
                         \def\Cpm{\widetilde \chi^\pm}
                         \def\Cmp{\widetilde \chi^\mp}
\def\l{\mbox{$ \lambda $}}
\def\lp{\mbox{$ \lambda' $}}
\def\lpp{\mbox{$ \lambda''$}}
                         \def\sq{\widetilde q}
                         \def\su{\widetilde u}
                         \def\sd{\widetilde d}
                         \def\sc{\widetilde c}
                         \def\ss{\widetilde s}
                         \def\st{\widetilde t}
                         \def\sb{\widetilde b}
\def\sl{\mbox{$\widetilde \ell$}}
\def\se{\mbox{$\widetilde e$}}
\def\snu{\mbox{$\widetilde \nu$}}
\def\smu{\mbox{$\widetilde \mu$}}
\def\stau{\mbox{\widetilde \tau}}
\def\tm{\widetilde m}
\def\ll{\mbox{$l_L$}}
\def\er{\mbox{$e_R$}}
\def\ql{\mbox{$q_L$}}
\def\ur{\mbox{$u_R$}}
\def\dr{\mbox{$d_R$}}
\def\cv{{\cal V}}
\def\lsim{\:\raisebox{-0.5ex}{$\stackrel{\textstyle<}{\sim}$}\:}
\def\gsim{\:\raisebox{-0.5ex}{$\stackrel{\textstyle>}{\sim}$}\:}
\thispagestyle{empty}
\begin{flushright}                               
                                                    IISc/CTS/10-99\\ 
						    PRL/TH/99-5 \\
						    UH-511-947-99 \\
\end{flushright}

\vskip 25pt

\begin{center}

{\LARGE\bf    
       Fermion Dipole Moments in Supersymmetric Models with      
          Explicitly Broken $R$-parity.}
\vskip 25pt
         R.M. Godbole$^1$, S. Pakvasa$^2$, S.D. Rindani$^3$ and X. Tata$^2$ 

\bigskip
1. Centre for Theoretical Studies, Indian Institute of Science, Bangalore,
560012, India. E-mail: rohini@cts.iisc.ernet.in\\
\vskip 25 pt
2.  Department of Physics and Astronomy, Univ. of Hawaii at Manoa, Honolulu, 
HI 96822, U.S.A. E-mails: tata@phys.hawaii.edu, pakvasa@phys.hawaii.edu
\\
\vskip 25 pt
3. Theory Group, Physical Research Laboratory, Navrangpura, Ahmedabad, 380 009,
India. Email: saurabh@prl.ernet.in.\\

\bigskip
           Abstract
\end{center}

\begin{quotation}
\noindent
We present a simple analysis that allows us to extract the 
leading mass dependence 
of the dipole moment of matter fermions that might be induced by new
physics. We present explicit results for the supersymmetric model
with broken $R$-parity as an illustration. 
We show that
the extra contributions 
to the electric dipole moment(edm)  of fermions from \rpv\ interactions 
can occur only at two loop level, contrary to claims in the literature.
We further find that
unlike the generic lepto-quark 
models, the extra contributions to the dipole moments  of the leptons can 
only be enhanced  by $m_b/m_l$ and not by $m_t/m_l$ relative to the
expectations in the Standard Model.
An interesting feature about this enhancement of 
these dipole moments is that it does not involve unknown mixing angles.
We then use experimental constraints 
on the electric dipole moments of $e^-$ and $n$ 
to obtain bounds on (the imaginary part of) products of \rpv\ couplings,
and show that bounds claimed in the literature are too stringent by many
orders of magnitude.

\end{quotation}
\newpage
\section{Introduction}
Dipole moments of the fermions have played a very  important role in
testing our understanding about particles and their interactions,
$(g-2)_\mu$ being  the first and foremost example. Indeed magnetic
and electric dipole 
moments, whether  diagonal or transition, 
have proved to be
excellent tools to constrain  extra physics beyond the Standard 
Model~\cite{electron,neutron}.  
The stringent experimental upper limits on lepton number violating
processes such as $\mu \rightarrow e\gamma$ as well as 
the small value of a possible
(Majorana) mass for neutrinos both lead to strong constraints on
transition moments of leptons that might be present in extensions of the
Standard Model (SM) that allow for lepton flavour violation.
In the SM the $CP$ odd neutron electric dipole moment (edm) vanishes at
two loops \cite{shabalin}. At three loops it has been
estimated~\cite{nshab,neutron} to be $d_n \sim 10^{-32 \pm
1}$~e~cm. Since there are no purely leptonic sources of CP violation in
the SM, an electron dipole moment can only be induced from $d_n$ at
second order in $G_F$ and thus may be estimated to be $d_e \sim (G_F
m_n^2)^2 d_n \sim 10^{-42} $~e~cm, to be compared to the estimate
$8\times 10^{-41}$~e~cm quoted in the literature
\cite{booth, electron}.  These expectations are much lower than the
current experimental limits~\cite{harris,commins}.  An observation of
the edm of either the electron or the neutron in current experiments
would clearly be very exciting since it would almost certainly signal
new physics. Many extensions of the SM, including the Minimal
Supersymmetric Standard Model (MSSM) where \rp\ is assumed to be
conserved include additional sources of $CP$ violation so that fermion
edms are induced at even the 1-loop level Such models are, of course,
severely constrained by the experimental bounds on neutron and electron
edms~\cite{masiero}.

Dipole moment operators flip chirality, and hence have either to be
proportional to {\it some} fermion mass (this may not be the mass of the
external fermion), or to a chirality flipping Yukawa type coupling
\cite{barrzee}.
The theoretical predictions for the moments of the heavier fermions like
the $t$, $b$ or the $\tau$, are larger than for first generation
particles due to the linear dependence on
$m_f$~\cite{hollik}. In models with lepto-quarks, particularly
large enhancements of the predicted values of the $\tau$ moments by a
factor of $m_t/m_\tau$, are possible~\cite{taulq}. Hence,
measurements of the dipole moments of the $\tau$ and the $t$, at
current~\cite{ackerstaff} or future $e^+e^-$ colliders~\cite{saur_pou} or
$\gamma\gamma$ colliders \cite{sp2}
form a potentially interesting probe of non-standard physics.
In this note, we set up 
a method of analysis, which would allow us to  extract the leading fermion mass
dependence of the coefficient of the induced dipole moment in any
theory.  We illustrate our method using the example of \rpv\
supersymmetric (SUSY) interactions~\cite{rpvint}.

Even assuming the MSSM field content, the most general renormalizable,
gauge invariant superpotential allows terms which do not conserve
$R$-parity. 
These terms also violate lepton number ($L$) and/or baryon number ($B$) 
conservation. Phenomenologically, many of
these \rpv\ couplings have been constrained very tightly using a large
number of low-energy and collider measurements~\cite{review-rpar}.
While $R$-parity violation does not appear to be required by any theoretical or
phenomenological considerations, it is not excluded either.  One of the
theoretical challenges then is to understand why some of the \rpv\
couplings in the superpotential are as small as they are (of course, we
have no understanding of why the electron Yukawa coupling is as small as
it is either), and especially, why products of $B$ and $L$ violating
couplings are so tiny. Many authors have examined phenomenological
implications of
$R$-parity violating models which differ in many ways from MSSM expectations. 

The \rpv\ part of the supersymmetric Lagrangian (assuming the MSSM field
content) is given by, 
\be 
{\cal L}_{\not\!\!R_p}=\left[ \l_{ijk} L_i L_j
E_k^c + \lp_{ijk} L_i Q_j D_k^c + \lpp_{ijk} U_i^c D_j^c D_k^c +
\kappa_i L_i H_2 \right]_F + {\rm h.c.} ,
\label{eq:eq1}
\ee
where $L_i, Q_i$, are the left handed lepton and quark $SU(2)$ 
doublet  superfields  corresponding to the three generations and 
$E_i^c, D_i^c, U_i^c$  are the $SU(2)$ singlet
lepton and quark superfields, with $i,j,...$ being the generation index,
while $H_2$ is the Higgs superfield with 
$Y = 1$.  
The first, second and the fourth terms violate lepton number conservation, 
while the third one violates conservation of $B$. 
The $\l_{ijk}$ and $\lpp_{ijk}$ are antisymmetric in the indices i,j and j,k 
respectively. The stability of the proton requires that both 
$B$ and $L$ violating operators  not be simultaneously present. 

Since these \rpv\ interactions violate L and/or B, their contributions
to masses and magnetic moments of the neutrinos, as well as the flavour
changing off diagonal moment $\mu \go\ e \gamma$ have been considered in
literature.  The dipole moment (direct or transition) of fermions can be
obtained by considering the matrix element of the electromagnetic or the
neutral weak current between on-shell fermions, as the momentum transfer
$q$ between these vanishes.  The (electromagnetic and weak) magnetic and
electric dipole moments are then given by the values of the tensor form
factors $F_T^\cv$ and $F^{'\cv}_T$ (at zero momentum transfer) that can
be read off as coefficients of the terms
$$\bar u_{f_1} (p-q)\sigma_{\mu \nu} q^\nu (F_T^\cv + \gamma_5
F^{'\cv}_T)u_{f_2}(p), \ (\cv=\gamma, Z)$$ in this matrix element. Of
course, in any renormalizable theory such as $R$-parity violating SUSY,
these coefficients which are only induced at the loop level must be
finite.  It has been claimed~\cite{frank} that {\it current} bounds on
the edm of the electron can be translated to very stringent bounds, {\it
e.g.}  $|\lambda'_{133}|^2 < 4 \times 10^{-10}$ if the additional $CP$
violating phases that are intrinsically present in these models are
large. This is traced to enhancement factors $\sim m_t/m_e \sim 10^5$
that are claimed to be present.

In this note, we develop a method that enables us to extract the
dependence of the coefficients $F_T^\cv$ and $F^{'\cv}_T$ on various
fermion masses in any extension of the SM. We then apply this to the
MSSM and to \rpv\ SUSY. In the first part of the 
analysis we consider only the trilinear \rpv\ terms and come back 
to the bilinears at the end. While our analysis of the mass
dependence of the induced quark and lepton dipole moments
yields answers consistent with many previous explicit
calculations, we find that the claimed enhancement of
the electron edm by the factor $m_t/m_e$ does not occur, and conclude
that the bounds on \rpv\ couplings coming from the edm and 
frequently quoted\cite{frank,adh,rev} in the literature are too stringent.

\section{Selection rules for non-zero dipole moments}
We begin by defining five different global charges $Q_{\ll},Q_{\er},Q_{\ql},
Q_{\dr}$ and $Q_{\ur}$ corresponding to five different $U(1)$ transformations. 
Only the 
members of the superfield indicated by the subscript have nonzero value of 
the particular charge, \eg ,
\newpage
\bearr
Q_{\ll} &=& 1\;\; {\rm for} \;\; e_{iL}, \nu_i,\widetilde e_{iL},
\widetilde \nu_i \;\;(i=1-3) \nonumber\\
&=& 0\;\; {\mbox{\rm for all the other particles/sparticles}}.
\label{eq:eq3}
\eearr 
Notice that {\it all} left-handed lepton fields and their superpartners
have the same charge, regardless of generation.
The other charges are similarly defined. The value of all the
charges for all gauge and Higgs bosons and their SUSY partners is zero. 

We see that gauge (and gaugino) interactions conserve these charges,
while superpotential Yukawa interactions and the soft SUSY breaking
$A$-terms as well as the \rpv\ terms in the Lagrangian, do not. These
charges are a kind of `super-chirality' in that they are non-zero even
for spin zero sfermions, as they must be in order to be compatible with
supersymmetry. They differentiate fermions of different chirality, and
also right handed quarks of different electrical charge. They do not,
however, differentiate between flavours of leptons or quarks with the
same chirality, and so are conserved by inter-generational
quark mixing.

The induced  dipole moment operator  will have to flip the 
chirality of the fermion involved as symbolically shown in 
\begin{figure}[htb]
\vspace{0.2cm}
\begin{center}
\mbox{\epsfig{file=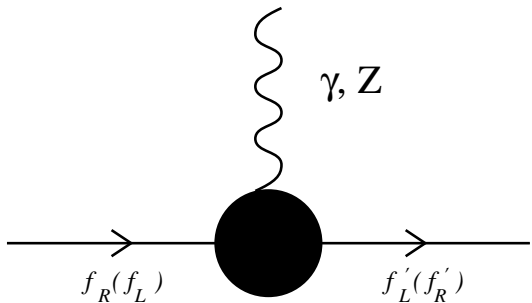, height=40mm}}
\caption{Generic diagram which will contribute to the dipole moment.}
\label{figone}
\end{center}
\vspace{-0.5cm}
\end{figure}
Fig.~\ref{figone}. In the SM, Yukawa interactions (we regard fermion
masses as Yukawa interactions) are the only source of chirality
flip. Within the MSSM, this is still the case if we include the
interactions of higgsinos as well as $A$-terms (proportional to the same
Yukawa coupling) in our definition of Yukawa
interactions.\footnote{Gaugino masses can flip the chirality of
gauginos, but in order for the chirality-flipped gaugino component to
couple, one also needs $\tilde{f}_L-\tilde{f}_R$ mixing which has its
origin in Yukawa interactions.} \rpv interactions are yet another source
of matter fermion chirality flip.

The charge assignments that we introduced in Eq.~\ref{eq:eq3} above
provide a systematic way of analysing when a dipole moment can be induced.
For instance, a leptonic moment requires $Q_{\ll}$ and $Q_{\er}$
to change by one unit in equal and opposite directions, with no change
in the other charges. Likewise, a moment for the $u(d)$ quark requires a
corresponding change in $Q_{\ql}$ and $Q_{\ur}$ ($Q_{\dr}$ and
$Q_{\ql}$). Since the change induced in each of these charges by any
(chirality-flipping) interaction is known (see Table 1), it is
straightforward to derive relations between the number of vertices of
various types of chirality flipping interactions in order that these
collectively induce a dipole moment for any particular matter
fermion.~\footnote{ These would only be necessary conditions since, without
further study, it cannot be guaranteed that the answer would not vanish.}

\begin{table}
\begin{center}
\caption{The change in the charges $Q_{\ll}, Q_{\er}, Q_{\ql},Q_{\ur}$
and $Q_{\dr}$ defined in the text for different interactions that might
be present in SUSY models with MSSM field content. Gauge and gaugino
interactions or Higgs and higgsino self interactions do not change any
of these charges.${\cal H}^0$ indicates any of the neutral Higgses in the MSSM.}
\bigskip
\begin{tabular}{|l|ccccc|}
\hline
Interaction & $\Delta Q_{\ll}$ & $\Delta Q_{\er}$ &$\Delta Q_{\ql}$  &
$\Delta Q_{\ur}$ & $\Delta Q_{\dr}$ \\
&&&&&\\
\hline
Lepton Yukawa Interactions & -1 & +1 & 0 & 0 & 0 \\
Up quark Yukawa Interactions & 0 & 0 & -1 & +1 & 0 \\
Down quark Yukawa Interactions & 0 & 0 & -1 & 0 & +1 \\
${\cal H}^0 H^- \tilde d^*_R \tilde u_R$, $ H^- \tilde d^*_R \tilde u_R$
& 0 & 0 & 0 & -1 & +1 \\
\hline
$\l_{ijk} L_i L_j E_k^c$ interactions & -2 & +1 & 0 & 0 & 0 \\
$\lp_{ijk} L_i Q_j D_k^c$ interactions & -1 & 0 & -1 & 0 & 1 \\
$\lpp_{ijk} U_i^c D_j^c D_k^c$ interactions & 0 & 0 & 0 & 1 & 2 \\
\hline
\end{tabular}
\label{tab:cases}
\end{center}
\end{table}

The changes in these charges for each of the vertices in the \rpv\ SUSY
model with MSSM field content is shown in Table 1. Of course, the
Hermitean conjugate of any interaction would lead to exactly the
opposite change in the charges. It should be
clarified that by Yukawa interactions, we mean all interactions with the
corresponding Yukawa coupling (with one exception listed in the fourth
row of Table 1 whose origin is clarified below):
for instance, lepton Yukawa interaction would include the (charged and
neutral) Higgs couplings to leptons, the lepton mass term, Higgs
slepton/sneutrino 
couplings from the superpotential, as well as 
the corresponding $A$-terms (assumed to be proportional to
the lepton Yukawa coupling) and left-right slepton mixing terms, and
likewise for the up (down) type Yukawa interactions. 
In addition to the Higgs sfermion couplings discussed above, there are
trilinear Higgs-sfermion interactions from both $D$-terms as
well as $F$-terms of the type,
\bearr
1&:& {\cal H}^0 (\tilde  u_L^* \tilde u_L + \tilde u_R^* \tilde u_R)\nonumber \\
2&:& {\cal H}^0 (\tilde  d_L^* \tilde d_L + \tilde d_R^* \tilde d_R)\nonumber \\
3&:& H^-\tilde  d_L^* \tilde u_L \nonumber \\
4&:& H^- \tilde  d_R^* \tilde u_R  
\label{eq:eq5p}
\eearr
where  ${\cal H}^0$ indicates any of the neutral Higgses in the MSSM.
Out of these, the first three are just like gauge interactions, as far
as our charges defined in Eq.~\ref{eq:eq3} are concerned  and hence will 
not affect the selection rules we will derive below. However, the fourth
one, 
which arises from the superpotential and is $\propto (m_u m_d / m_W) 
\tilde d^*_R \tilde u_R$, violates the charges we have defined in 
Eq.~\ref{eq:eq3}, but changes them in a way different than the Yukawa 
interactions. This happens, even though
the term arises from the superpotential,  because it comes from a
cross term between a down-type Yukawa interaction and the (Hermitean conjugate)
of the up-type Yukawa interaction.   Note that this kind of vertex can exist 
only  for the squarks and not for the sleptons as there is no $\tilde \nu_R$
in the MSSM.  
There are also quartic scalar couplings between a pair of Higgses and 
$\tilde f_R^* \tilde f_R$ / $\tilde f_L^*\tilde f_L$ pairs.
Of these, only the
term ${\cal H}^0 H^- \tilde d_R^* \tilde u_R$, causes a nonzero charge 
change which is different from those caused by the Yukawa interactions. 
The changes in the charges for these interactions are given in the
fourth row of Table 1, and are as expected, the difference of the changes for 
down and up type Yukawa interactions.

We are now ready to compute the change in each of these charges for any graph.
Let us denote by $P,S$ and $R$ the number of down-quark, up-quark and
lepton Yukawa interactions (in the generalized sense explained above), 
and by $P^*,S^*,R^*$  
the  number of insertions corresponding to the Hermitean conjugate
($h.c.$) of these interactions. Similarly, let 
$N,M,L$  denote  the number of vertices 
corresponding to interactions proportional to $\l , \lp , \lpp$ of  
Eq.~\ref{eq:eq1} respectively, again with $N^*,M^*$ and $L^*$ indicating the
number of vertices corresponding to the $h.c.$ of these
interactions. Finally, let $T (T^*)$ denote the number of trilinear or
quartic scalar vertices corresponding to the interactions in the fourth
row of Table 1.

It is easy to see from Table 1 that the net change in various charges,
is given by,
\bearr
\Delta Q_{\ll} & = & -2 \Delta N - \Delta M -\Delta R \nonumber \\
\Delta Q_{\er} & = & \Delta N + \Delta R \nonumber\\
\Delta Q_{\ql} & = & -\Delta M -\Delta P -\Delta S \nonumber \\
\Delta Q_{\dr} & = & 2\Delta L + \Delta P  + \Delta M + \Delta T \nonumber \\
\Delta Q_{\ur} & = & \Delta L + \Delta S - \Delta T  ,
\label{eq:eq6}
\eearr
where $\Delta M$, is given by $\Delta M = M - M^*$, {\it etc}. 
Now we can solve this general system of equations
for the special cases of the moments of the leptons as well as the up/down 
quarks. 

\noindent 
\underline{Leptonic Moments}:
Let us now consider the case where $f,f'$ of 
in Fig.~\ref{figone} are leptons.
In this case we must have, 
\be
\Delta Q_{\ll} = -1, \Delta Q_{\er} = 1,
\label{eq:eq7}
\ee
or vice versa  and all the other remaining charges should remain unchanged, \ie
\be
\Delta Q_{\ql} = 0, \Delta Q_{\ur} = 0, \Delta Q_{\dr} = 0.
\label{eq:eq8}
\ee 
Note that our analysis does not distinguish between direct and
transition moments. In this case using Eqs.~\ref{eq:eq6} we get, 
\bearr
\Delta N & = & 1- \Delta R \nonumber \\ 
\Delta M & = & \Delta R -1 \nonumber \\ 
\Delta P & = & 1- \Delta R -\Delta T \nonumber \\ 
\Delta L &=&0, \Delta S= \Delta T.
\label{eq:eq9}
\eearr
\noindent 
It is clear that any dipole moment ${\cal D}_l$ that this diagram can 
give rise to will be 
$$
{\cal D}_{l} \propto m_{l_i}^{R+R^*} m_{d_j}^{P+P^*}m_{u_k}^{S+S^*}
(m_{u_l}m_{d_l})^{T+T^*}
$$
with an appropriate numbers of the large masses (at least $M_W$ or
$M_{SUSY}$ depending on the graph) coming from the loops in the
denominator to give the right dimension.  Here, $m_{l_i}, m_{u_k}$ and
$m_{d_j}$ denote {\it some} lepton, up type quark and down type quark
mass.  We first see that if there are no $R$-parity violating
interactions (so that $\Delta L = \Delta M = \Delta N =0$), $\Delta R
=1$, so that at least $R$ or $R^*$ must be non-zero. In the MSSM (or the
SM) the leptonic dipole moment must, therefore, be proportional to some
lepton mass.\footnote{This assumes that the leptonic $A$-terms are
proportional to the lepton Yukawa coupling, which need not be the
case.}  Since there are no sources of lepton flavour violation, this
must be the external lepton mass. In \rpv\ models, the third of
Eqs.~\ref{eq:eq9} implies that no moment is possible without at least
one Yukawa interaction insertion corresponding to a lepton or a
down-type quark.
Due to the lepton number violation inherent in the \rpv\ interactions,
the fermion can be a $b$ quark and hence in principle, an enhancement of
the loop contribution to the moments by a factor of $m_b/m_l$ (relative
to the case with the SM/MSSM) is possible. The last of these equations
tells us further that up-type quark masses enter only as even powers so
that these can never be the sole source of the required chirality flip
for a lepton dipole moment.  Indeed these masses have to be {\it in
addition to} the lepton or down type mass as mentioned above, and so
will necessarily be accompanied by the same power of some high mass in
the denominator, and so will actually suppress the moment. Clearly
the claims \cite{frank,adh,rev} that the edm of the electron may be
enhanced by factors of $m_t/m_e$ in \rpv\ models do not seem tenable.
\footnote{We may also add here that the analysis in Ref.~\cite{frank}
appears to be based on the computation of one-loop diagrams that do not
exist.}

We mention in passing that the conditions of Eqs.~\ref{eq:eq9} that we
have derived have also to be satisfied by the diagrams that lead to the
Majorana mass~\cite{mass,anjan-vempati,biswa} or 
dipole moments of the neutrinos~\cite{moment}, as well as 
$\mu \go e \gamma$ in \rpv\ theories \cite{many}.

\noindent
\underline{Down-type quark moments}:
For the case of the down quark moments we have to solve the system of equations
\be
\Delta Q_{\ql} = -1, \Delta Q_{\dr} = 1,
\label{eq:eq10}
\ee
with all the other charges remaining unchanged, \ie
\be
\Delta Q_{\ll} = 0, \Delta Q_{\ur} = 0, \Delta Q_{\er} = 0.
\label{eq:eq11}
\ee
From eq.~\ref{eq:eq6} we find,
\bearr
\Delta M & = & 1- \Delta P - \Delta T \nonumber\\
\Delta N & = & \Delta P -1 + \Delta T \nonumber\\
\Delta R & = & 1- \Delta P  -\Delta T \nonumber \\
\Delta L &=& 0,\Delta S=\Delta T.     
\label{eq:eq12}
\eearr
Once again we first see that if there are no \rpv\ interactions, the dipole
moment would vanish in the absence of all down-type Yukawa couplings. 
Again the \rpv\ contributions to the dipole 
moments of down-type quarks are nonzero only if either $\Delta R$ or 
$\Delta P$ are non-zero,
and thus are proportional either to a lepton mass or a down-type quark 
mass. Thus 
for a $d$-quark, for example, enhancements of $m_b/m_d$ are possible. Any 
dependence on $m_t$ will come only in even powers and suppressed by heavier 
masses in the denominator compared to this leading order mass
dependence, and possibly also by
small Kobayashi-Maskawa (KM) matrix elements.  Thus no big enhancements of the \rpv\ contributions 
to the dipole moment due to the large top mass are possible.

\noindent
\underline{Up-type quark moments}:
For this case, we need 
\be
\Delta Q_{\ql} = -1, \Delta Q_{\ur} = 1,
\label{eq:eq13}
\ee
with the other charges remaining unchanged, \ie
\be
\Delta Q_{\ll} = 0, \Delta Q_{\dr} = 0, \Delta Q_{\er} = 0.
\label{eq:eq14}
\ee
Using Eqs.~\ref{eq:eq6} we get,
\bearr
\Delta R & = & -\Delta N \nonumber\\
\Delta P & = & \Delta N - \Delta T \nonumber\\
\Delta M & = & - \Delta N \nonumber \\
\Delta L &=& 0,\Delta S=1 + \Delta T.
\label{eq:eq15}
\eearr 
In this case a solution without an up-type Yukawa interaction is
not allowed as opposed to the earlier two cases where a solution was
allowed where a single power of quark (lepton) mass could appear for the
lepton (quark) moment. The leading mass dependence of an up-quark
moment generated by \rpv\ interactions is necessarily an up-type
mass.This happens because neither the \l\ or the \lp\ interactions
involve a $\tilde u_R$ or $u_R$. 

We also see that for contributions that
will involve only the \lpp\ part of the \rpv\ interactions, the dipole
moment for the down quark will thus be proportional to $m_{d_i}
m_{u_i}^{2n}$ as opposed to the up quark moment which will be proportional
to $m_{u_i} m_{d_i}^{2n}$ ($n=0,1,2...$). This is in agreement with the
result for the edm due to the \lpp\ couplings that was derived long
ago~\cite{barb}. 

A few more general comments are in order here. A natural question to ask is
why is it possible to get an enhancement of the dipole moments by a factor
of $m_t/m_f$ in the case of  theories with general 
lepto-quarks~\cite{taulq}. This can be traced back to 
the fact that in SUSY with \rpv\, even though the squarks/sleptons 
do play the role of lepto-quarks which have \lv\ or  \bv\
interactions, their couplings are chiral as a result of the supersymmetry, 
which allowed us the charge assignment made in eqs.~\ref{eq:eq3} in the 
first place.
The chiral nature of the couplings, therefore, forbids the enhancement of 
dipole moments of the leptons and down-type quarks  as compared to the
expectations in SM/MSSM, by a  factor of $m_t/m_l$ or $m_t/m_d$.

\section{Numerical estimates of electric dipole moments of the electron
and neutron}

We have discussed necessary conditions for any diagram in the MSSM or
\rpv\ framework to contribute to a lepton or quark dipole moment.
Of course, in order to conclude that the induced moment is an electric
dipole moment (weak or electromagnetic), one has to further check that
it has a $CP$ violating piece.  With the usual convention for phases of
the fields, if the amplitude in Fig.~\ref{figone} is complex we would
ensure that the induced edm is nonzero. We will see below that even for
arbitrary phases in the \rpv\ couplings, the dominant contributions to
the edm from \rpv\ couplings come only at two-loop level.

Let us start with the case of a lepton. The lowest order diagrams
involving \rpv\ couplings will need
$N=N^*=1$ or $M=M^*=1$. It is then easy to check that within our
framework, each of these diagrams is proportional to some $|\lambda|^2$ 
($|\lambda'|^2$) and so cannot contain a complex piece from the \rpv\
coupling. 

\begin{figure}[htb]
\begin{center}
\mbox{\epsfig{file=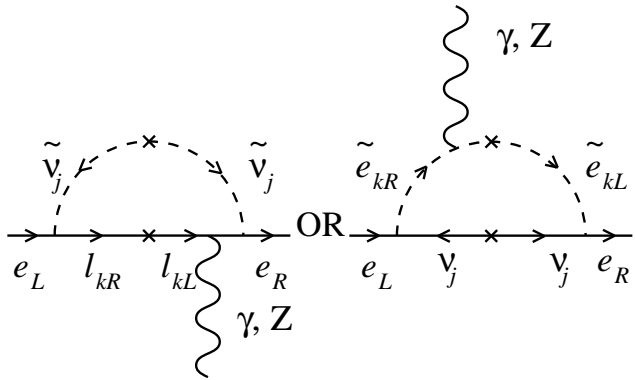, height=50mm}}
\caption{One loop diagrams contributing to the edm of electron in the presence
of Majorana masses for the neutrinos(sneutrinos). Note that here and in all 
the other diagrams, the $\gamma / Z$ is to be attached to all possible lines 
to which it couples.}
\label{figtwo}
\end{center}
\vspace{-.5cm}
\end{figure}

We note in passing that if the model is extended to allow for
Majorana masses for neutrinos (and lepton number violating sneutrino
masses to preserve supersymmetry), an electron edm would be possible
from diagrams shown in Fig.~\ref{figtwo}. 
The `cross' on neutrino line (or the corresponding
sneutrino line ) essentially corresponds to this `Majorana' mass
insertion for neutrinos (lepton number breaking sneutrino mass
insertion), which is not present in the MSSM.\footnote{The selection
rules above would, of course, then have to be modified to allow for
these additional couplings.} The `cross' on the charged lepton line (the
charged scalar lepton) corresponds to a mass insertion (the insertion of
the $L-R$ slepton mixing). We see that each of these diagrams is
proportional to products of $\lambda$'s (not $\lambda$ times $\lambda^*$)
and so can lead to an edm for the electron.  We expect though that the 
contribution will be extremely small due to the smallness of the 
neutrino masses. The corresponding contribution from a (SUSY violating)
sneutrino mass insertion may be worthy of examination. The same
mechanism is clearly not possible with $\lambda'$ type couplings as
there is no neutral particle in the loop.

Once we go to two loops, it is simple to see that there are many types of 
diagrams involving \rpv\ couplings, where 
product of the relevant couplings is complex. An example is shown in
\begin{figure}[htb]
\begin{center}
\mbox{\epsfig{file=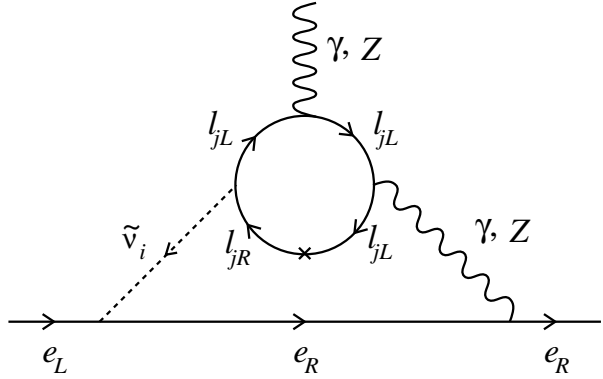, height=50mm}}
\caption{An example of the  leading two-loop contribution to  the edm 
of electron due to \l\  couplings.}
\label{fig3}
\end{center}
\vspace{-0.5cm}
\end{figure}
Fig.~\ref{fig3}. This corresponds to the case $N=N^*=1 $. According to
our general analysis therefore, the contribution to the dipole moment
should be proportional to some $m_{l_j}$ which need not be the mass of
the external lepton (electron in this case). 
This amplitude involves two {\it different} \l\ couplings and hence is
complex in general. Here the source of the complex nature of the
amplitude and hence for the edm, is the irremovable phases of the \rpv\
couplings.  We estimate the order of magnitude of 
the real part of the edm as the product of
explicit factors of couplings, mass insertions and colour factors, a
factor of $1/(4\pi^2)$ for each loop, and finally appropriate powers of
the ``large mass'' ($m_{\tilde \nu}$ in this case) in the denominator to
get the appropriate dimension. We then take the edm to be the imaginary
part ($\Im$) of this product.\footnote{We
tacitly assume that the \rpv\ couplings that we have written are in the
mass eigenstate basis for matter fermions and sfermions. The mass
insertions in the figures are merely to show explicit fermion mass
factors that would arise from the computation.  In other bases, there
could be complex contributions from off-diagonal terms in the
propogators, and it would be difficult to isolate the imaginary part of
the contribution. Since supersymmetry is broken by sfermion mass terms,
SUSY would appear to be explicitly broken by some interactions such as
trilinear scalar couplings. But this is not relevant to our analysis as
we include soft SUSY breaking couplings anyway. The important thing is
that our selection rules of the previous section are not affected by
this. We stress that it is only for the extraction of the imaginary
part, and {\it not for the derivation of the selection rules} are we
forced to go into this basis.}  We will, of course, overestimate the
answer in the event there are significant cancellations between several
diagrams. For the diagram in Fig.~\ref{fig3} we obtain,
\be d_e \sim {{(e^2,g_Z^2)} \over {4 \pi^2}} {1 \over {4 \pi^2}} \Im
\left[\sum_{ij,i\ne 1,j} m_{l_j} \lambda_{ijj}^{*} \lambda_{i11} {1
\over {m_{\tilde \nu _i}^2}}\right].
\label{eq:delam}
\ee 
As long as $|\lambda_{233}|$ is not unduly small, the dominant
contribution will be the one corresponding to $j=3$.  Due to the
antisymmetry of the \l\ couplings in the first two indices this piece is
then given by 
\be d_e \sim {{(e^2,g_Z^2)} \over {4 \pi^2}} 
{1 \over {4 \pi^2}} 
{m_\tau \over {{m_{\tilde \nu_2}^2}}} 
\Im \left(\l_{211} \l_{233}^*\right)
\label{eq:delam1}
\ee
and hence we see that this diagram gives an enhancement of the dipole moment
by a factor ${m_{\tau}/ m_e}$. The interesting feature of this
enahancement is that the large mass has appeared without paying any price 
for the mixing angles.

At two loop level one can also get a diagram corresponding to $\Delta M
= -1, \Delta N= 1,\Delta P=1$ but with $\Delta R=0$, which gives an
enhancement of the edm of the electron by a factor of ${m_b/m_e}$ as
we have already discussed. An example is
shown
\begin{figure}[htb]
\begin{center}
\mbox{\epsfig{file=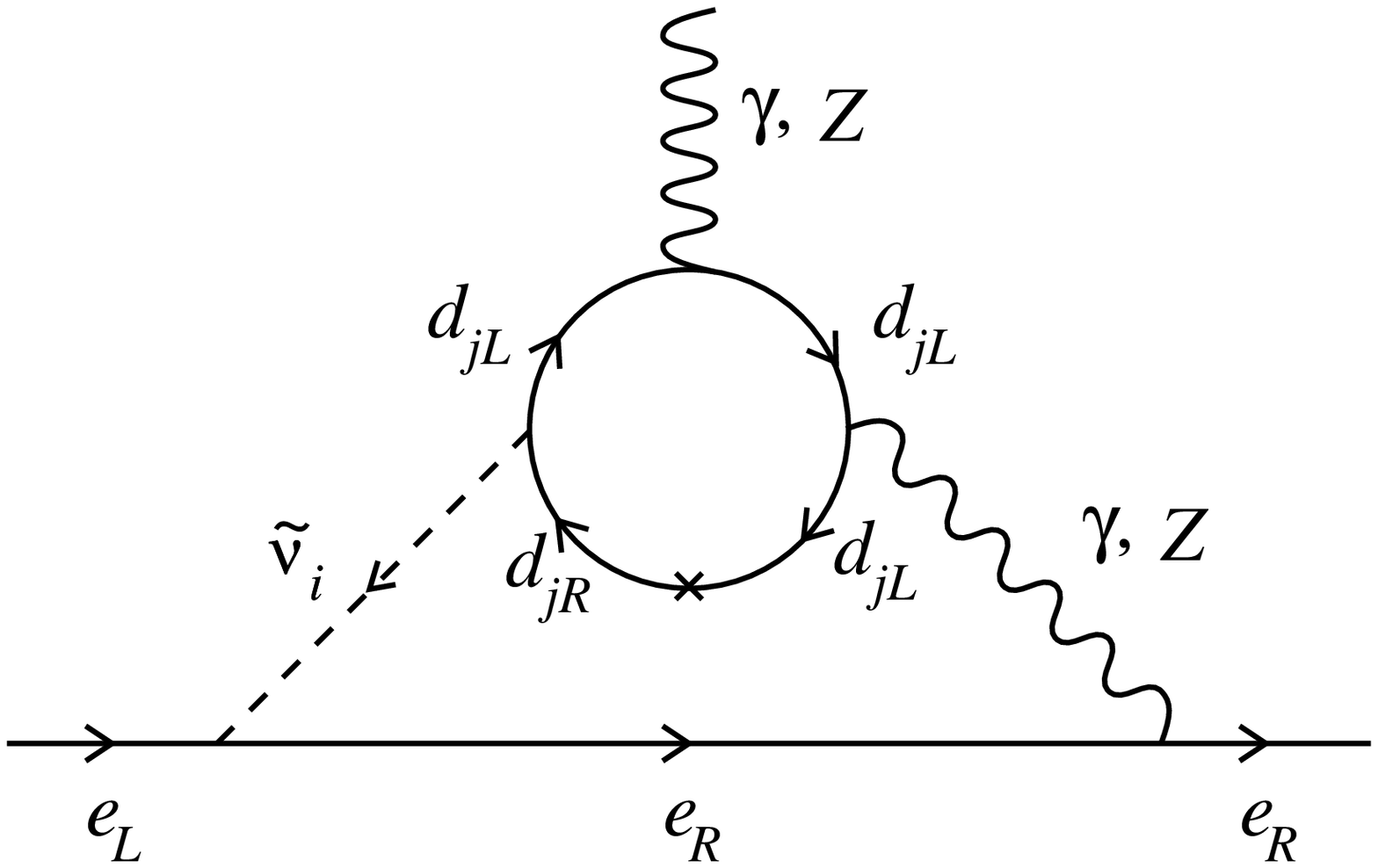, height=50mm}}
\caption{An example of the leading two loop contribution to the electron edm 
due to \lp\ and \l\  couplings.}
\label{fig4}
\end{center}
\vspace{-0.5cm}
\end{figure}
in Fig.~\ref{fig4}. Again the corresponding contribution is  
similar to one in  Eq.~\ref{eq:delam} and is dominated by the $j=3$ 
term. The dominant piece is estimated by,
\be
d_e  \sim {{(e^2,g_Z^2)} \over {4 \pi^2}} {1 \over {4 \pi^2}}  m_b
\Im \left[\sum_{i\ne 1} 3 \lambda_{i33}^{'*} \lambda_{i11} 
{1 \over {m_{\tilde \nu_i}^2}}\right],
\label{eq:dellp}
\ee
where a colour factor of 3 has been inserted.
It should be noted here that  in both the cases there are additional 
diagrams where the \snu\ and  the neutral gauge boson lines are 
crossed, as well as where the $\snu_{L}$ is replaced by $\se_{L}$ and 
$\gamma /Z$ replaced by $W$. 
They would give contributions similar to the one given above, except that
the $m_{\snu _{i}}^2$ will be replaced by $m_{\se_{iL}}^2$. There exist 
a large 
number of two loop contributions involving the Higgs exchanges, but in all 
the cases the resulting contributions are proportional to 
$m_e$ or even higher powers in agreement with the expectations from 
our general rules.

We can use these estimates to constrain products of \l\
couplings or those of \l\ and \lp\ couplings using the current experimental
limits on the edm of the electron. 
The potentially largest contributions, and hence the best limit will
come from the contributions of the diagrams shown in Figs.~\ref{fig3}
and ~\ref{fig4}. Using Eqs.~\ref{eq:delam1} and ~\ref{eq:dellp} and the
current bound~\cite{harris} $d_e < 10^{-27}$~e~cm on the edm of the $e$,
we get 
\bearr \Im \left( \l_{211} \l_{233}^* \right ) &<& 5 \times
10^{-4} \left({{\tilde m}\over{1TeV}}\right)^2 \nonumber\\ 
\Im
\sum_{i \ne 1} \l_{i11} \lp_{i33}^*  & < & 0.6 \times 10^{-4}
\left({{\tilde m} \over {1 TeV}}\right)^2
\label{eq:constr_el}
\eearr 
Here $\tilde m$ stands for the mass of the appropriate SUSY
scalar. The improvement in the second case is simply the factor of
$m_b/m_\tau$ and the colour factor. We stress that our estimates are
crude: we have clearly not made a complete computation of any amplitude,
and also not included contributions from other diagrams. We add
though that for some cases where explicit computations~\cite{darwin} are
available in the literature, we did check that our crude estimate gives
reasonable agreement with the complete calculation.

In case of the $d$-quark dipole moment, again there is a counterpart of
the diagram in  Fig.~\ref{fig3} where \l\ couplings will be replaced by \lp . 
The diagram is similar to the ones shown in Figs.~\ref{fig3},~\ref{fig4}.
It is obtained by replacing in Fig.~\ref{fig3} $e_L,e_R$ by $d_L,d_R$ and 
the leptons in the central loop by quarks. In this the dominant contribution 
is obtained from the $b$ quark in the loop.
The \lp\ couplings have no antisymmetry  and the dominant contribution in 
this case, is estimated by,
\be 
d_d  \sim {{(e^2,g_Z^2)} \over {4 \pi^2}} {1 \over {4 \pi^2}} m_b
\Im \left[ \sum_{i}  3 \lambda_{i33}^{'*} \lambda_{i11}^{'} {1 \over {m_{\tilde \nu_{i}^2}}}\right].
\label{eq:ddlpZ}
\ee
There exist possible two loop diagrams with charged Higgs exchange and \lp\ 
couplings. An example is shown in  
\begin{figure}[htb]
\begin{center}
\mbox{\epsfig{file=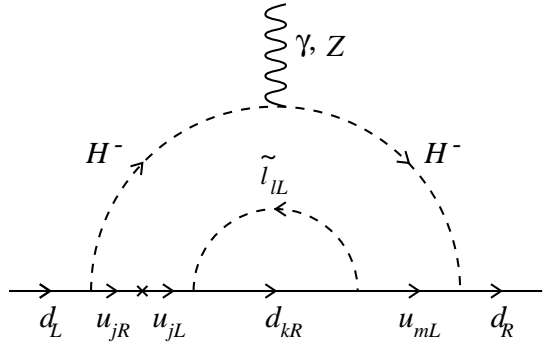, height=45mm}}
\caption{An example of the Higgs mediated  two loop contribution to 
the edm of the down quark due to \lp\ couplings.}
\label{fig5}
\end{center}
\vspace{-.5cm}
\end{figure}
Fig.~\ref{fig5}. Here the dominant contribution is again given by $j=3$ and
can be estimated to be,
\be 
d_d  \sim {{(e^2,g_Z^2)} \over {4 \pi^2}} {1 \over {4 \pi^2}} 
{{m_t^2 m_d } \over {m_W^2}} {1 \over M^2}
\Im \left[\sum_{l,k} \lambda_{l3k}^{'*} \lambda_{l1k}^{'} V_{td}^*
V_{ud} \right ], 
\label{ddlph}
\ee 
where $M = \max({m_{\tilde e_L}, {m_{H^-}}})$, and $V_{ud}$ etc. are
the elements of the KM matrix.
We see that
this is proportional to $m_t^2 m_d/m_W^2$
In addition there is suppression due to the small KM mixing
angles as well. Hence, the contribution of the diagram shown in
Fig.~\ref{fig4} and the related diagrams, proportional to $m_b$, is
still the dominant one in spite of the $m_t^2$ factor here. This is an
illustration of our general statement that the dipole moment for the
down-type quark (and also of the lepton) does not receive enhancements
due to the large top mass.

The edm of the d-quark due to the \lpp\ couplings had been estimated 
previously~\cite{barb}  and the corresponding diagram is shown in
\begin{figure}[htb]
\begin{center}
\mbox{\epsfig{file=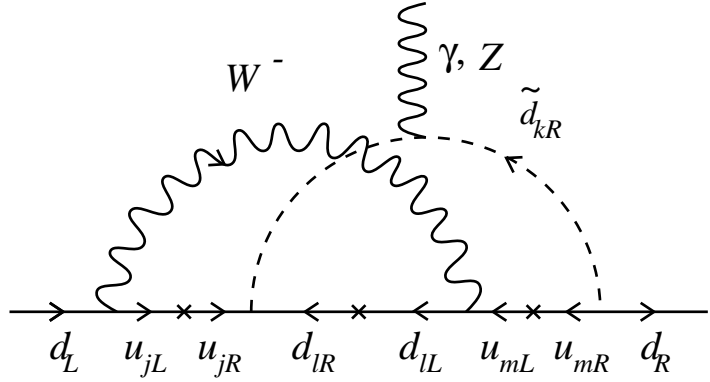, height=50mm}}
\caption{An example of the leading two loop contribution to the down quark edm 
due to \lpp\ couplings.}
\label{fig6}
\end{center}
\vspace{-.5cm}
\end{figure}
Fig.~\ref{fig6}. Here the dominant contribution is again given by $j=m=3$. The
antisymmetry of the \lpp\ couplings ensures that the dominant contribution
is then given by, 
\be 
d_d  \sim {{(e^2,g_Z^2)} \over {4 \pi^2}} {2 \over {4 \pi^2}} 
{{m_t^2 m_b}  \over {m_W^2}} {1 \over {m_{\tilde {bL}}^2}}
\Im  \left[ V_{td}^*V_{tb} \lambda_{323}^{''} \lambda_{321}^{''*} \right].
\label{eq:ddlpp}
\ee
Due to the relative values of the KM matrix elements $V_{td}$ and
$V_{cd}$, the contribution corresponding to $j=2,m=3$ is down compared
to the one above by about a factor 5, assuming comparable \rpv\
couplings in the two cases. Even this contribution
corresponding to $j=3,m=3$, is still small compared to that of
Eq.~\ref{eq:ddlpZ} for similar values of the \lp\ and \lpp\ couplings,
due to the small KM matrix elements. The diagram obtained from
Fig.~\ref{fig6}, by replacing $W^-$ by $H^-$ also makes a similar
contribution. In this case, the `crosses' corresponding to the mass
insertions on the $u_j$ and $u_m$ lines in the diagram are not present
as the chirality flip is achieved at the two $H^-$ vertices. This will
contain an additional factor of $\cot^2 \beta$, where $\tan \beta$ is
the ratio of the vacuum expectation values ($vev$s) of the $Y=1$ and
$Y=-1$, neutral Higgs fields.

\begin{figure}[htb]
\begin{center}
\mbox{\epsfig{file=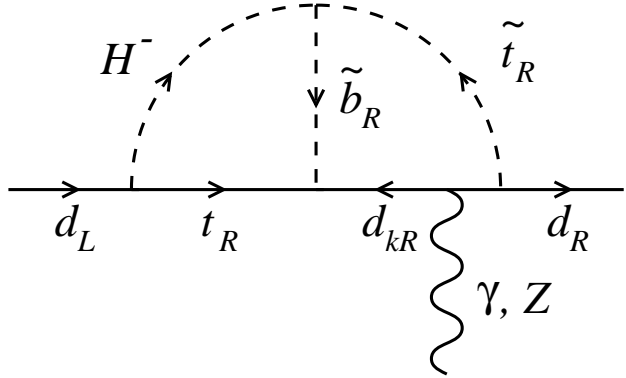, height=50mm}}
\caption{An example of the  two loop contribution to the edm of the down 
quark due to \lpp\ couplings and supersymmetric interactions of the Higgs.}
\label{fig7}
\end{center}
\vspace{-.5cm}
\end{figure}

At this point it is in order to look at potential contributions to the edm of 
$d$ from diagrams involving the scalar couplings in the fourth row of
the Table 1. There are no analogous contributions to lepton moments. An
example is shown in Fig.~\ref{fig7}.
The contribution of this diagram is given by, \be d_d \sim {g^2 \over {4
\pi^2}} {2 \over {4 \pi^2}} { m_t^2 m_b \over {m_W^2}} {1 \over {M^2}}
\cot\beta\ \Im
\left[ 
V_{td}^*{\tilde U}_{tb}^R \lambda_{323}^{''}\lambda_{321}^{''*}
\right], 
\label{eq:row4}
\ee 
with $M =
\max({m_{\tilde q}, m_{H^-}})$, and $g$ the $SU(2)$ gauge
coupling. Since the term originates from the superpotential it is not
surprising that the the contribution is given by an expression similar
to Eq.~\ref{eq:ddlpp}, with $V_{tb}$ replaced by $\tilde U_{tb}^R$, the
mixing matrix element in the right-squark sector. Note however, the edm
is nonzero even with no intergenerational mixing in the squark sector.
It is amusing to note that there is also a contribution to the edm of
the $d$-quark if there is no inter-generational mixing even in the quark
sector -- simply replace the $t$-quark by a $u$-quark. This contribution
is smaller than that in Fig.~\ref{fig7} by a factor
$m_u/|V_{td}|m_t$.

The up-quark moment will receive contributions from the \rpv\
interactions from diagrams similar to those shown in Figs.~\ref{fig5}
and \ref{fig6}, by simply interchanging $u$ and $d$ and changing $\tan
\beta$ by $\cot \beta$.  There does not exist a counterpart of the
diagrams shown in Figs.~\ref{fig3},\ref{fig4}, for the $u$-type quark.
The dominant contribution will be given by an expression similar to that
of Eq.~\ref{eq:ddlpp} involving different elements of the KM matrix and
again the roles of $m_{u_i}$ and $m_{d_i}$ reversed. Hence the dominant
piece is now sum of two terms which are proportional to $m_t m_b^2
V_{ub}$ and $m_t m_b m_s V_{us}$, in accordance with the general results
obtained from Eq.~\ref{eq:eq15}.  Both these terms are comparable to
each other in size because of the relative size of different elements of
the KM matrix.  The contributions due to diagrams involving charged
Higgs will involve a further factor of $\tan^2 \beta$. Since we have
just one factor of $m_t$, this contribution is suppressed relative to
that of Eq.~\ref{eq:ddlpp} by a factor 40 or so, which in turn is
smaller than that of Eq.~\ref{eq:ddlpZ} by about a factor 5. Hence
while estimating the $n$ edm, we will neglect the $u$ contribution
completely.

Using  then the current experimental result $d_n < 6.0 \times 10^{-26} $~e~cm,
and Eq.~\ref{eq:ddlpZ}, we get
\be
\Im \left [ \sum_{k} \lp_{k11} \lp_{k33}^* \right ] < 10^{-2}
\left({{\tilde m} \over {1 TeV}}\right)^2.
\label{constr_n_zee}
\ee 
Eq.~\ref{eq:ddlpp} yields a  weaker constraint: 
\be \Im
\left [ \lpp^*_{312} \lpp_{332} \right ] < 0.03 \times {{0.01} \over
{V_{td}}} \left({{\tilde m} \over {1 TeV}}\right)^2 
\ee 
The
contribution in Eq.~\ref{eq:row4}, as well as from other diagrams
involving a charged Higgs boson in the loop, leads to a similar bound.
In obtaining
these we have assumed that $d_n \sim d_d$. In view of the fact that we
have not really computed any of the diagrams, but
merely estimated the various contributions, it did not seem reasonable
to attempt to include the long distance contributions which could only
strengthen these bounds. 

We briefly mention the possibility of using the edm of heavier fermions,
in particular the tau and the top to constrain \rpv\ couplings.
As far as $d_{\tau}$ is concerned, the counterpart of the diagram of
Fig.~\ref{fig3} (as well as the corresponding diagrams with $W$) will
not contribute as the dominant piece proportional to $m_\tau$ will have
no imaginary part.  The diagram analogous to Fig.~\ref{fig4}, however,
does contribute to $d_{\tau}$. The real and imaginary parts of the {\it
weak} dipole moment of the $\tau$ have recently been constrained by the
OPAL Collaboration~\cite{ackerstaff} to be smaller than $6\times
10^{-18}$~e~cm and $1.5\times 10^{-17}$~e~cm, respectively. Clearly,
these limits do not give any significant constraints on the
corresponding \rpv\ couplings. There are no data on $d_t$ at this time.

Our analysis up to now has ignored
the bilinear terms in the superpotential of Eq.~\ref{eq:eq1}, and also
corresponding soft SUSY breaking scalar bilinears in the potential. 
In the case of exact SUSY, the former can be rotated away~\cite{rpvint} 
and, of course, the latter are absent, {\it i.e.} $R$-parity violation
occurs only through the trilinear interactions that we have analysed.
This is not, however, true in the realistic case where 
supersymmtery is broken, because the bilinear soft terms in the scalar
potential cannot simultaneously be rotated away.
Even if we assume that these are absent at some very high scale, these
terms, which are an additional source to the changes of the super-chiral
charges in Table \ref{tab:cases}, are generated~\cite{anjan,anjan-vempati}, 
by radiative corrections.  
A more important difference, however, is that in the presence of the scalar
bilinears the sneutrino fields generically acquire a $vev$, so that the
charge $Q_{l_L}$ is now no longer conserved. In principle, it would be
possible to include modifications to our analysis by allowing diagrams
where sneutrino fields disappear or are created from the vacuum: but the
result then depends on the number of fields that disappear into, or are
created from, the vacuum and the simple predictions that we have
obtained are lost. In models where bilinears are only radiatively
generated, sneutrino $vevs$ are smaller than a few GeV, and these
contributions are small, and our analysis yields a reasonable
approximation. There are also models~\cite{tonnis} where due to
additional discrete symmetries, sneutrino $vevs$ are absent. We have,
however, not analysed the generic case where the bilinear mass terms and
sneutrino $vevs$ are all of the order of the SUSY breaking scale. It
would be interesting to investigate whether this situation yields a new
possibility of generating large electric dipole moments for matter
fermions.

{\bf Acknowledgements:} R.M.G acknowledges the National Science
Foundation's U.S. India Cooperative Exchange Program under the
NSF grant INT-9602567 that enabled her to visit the High Energy Theory 
Group at the University of Hawaii where this work was initiated. S.D.R. 
thanks the Centre for Theoretical Studies, Indian Institute of Science, 
Bangalore, for hospitality during his visit. He also thanks Anjan Joshipura 
for discussions. This research was supported in part by
the US Department of Energy grant DE-FG-03-94ER40833.

\end{document}